\documentclass[twocolumn]{aastex631}

\usepackage{units}

\usepackage{amsmath}

\newcommand{\code}[1]{\texttt{#1}}
\newcommand{\mesa}{\code{MESA}}
\newcommand{\MESA}{\mesa}

\newcommand{\Msun}{\ensuremath{\mathrm{M}_\odot}}

\newcommand{\Teff}{\ensuremath{T_{\rm eff}}}

\newcommand{\nuclei}[2]{\ensuremath{\mathrm{^{#1}#2}}}

\newcommand{\carbon}[1][12]{\nuclei{#1}{C}}
\newcommand{\nitrogen}[1][14]{\nuclei{#1}{N}}
\newcommand{\oxygen}[1][16]{\nuclei{#1}{O}}
\newcommand{\fluorine}[1][19]{\nuclei{#1}{F}}
\newcommand{\neon}[1][20]{\nuclei{#1}{Ne}}

\begin{document}

\author[0000-0002-4870-8855]{Josiah Schwab}
\affiliation{Department of Astronomy and Astrophysics, University of California, Santa Cruz, CA 95064, USA}
\correspondingauthor{Josiah Schwab}
\email{jwschwab@ucsc.edu}

\author[0000-0002-4791-6724]{Evan B. Bauer}
\affiliation{Center for Astrophysics | Harvard \& Smithsonian, 60 Garden St Cambridge, MA 02138, USA}

\received{02 Jul 2021}
\revised{03 Aug 2021}
\accepted{05 Aug 2021}

\title{The final fates of close hot subdwarf - white dwarf binaries: mergers involving He/C/O white dwarfs and the formation of unusual giant stars with C/O-dominated envelopes}

\begin{abstract}
  Recently, a class of Roche-lobe-filling binary systems consisting of
  hot subdwarf stars and white dwarfs with sub-hour periods has been
  discovered.  At present, the hot subdwarf is in a shell He burning
  phase and is transferring some of its remaining thin H envelope to
  its white dwarf companion.  As the evolution of the hot subdwarf
  continues, it is expected to detach, leaving behind a low mass C/O
  core WD secondary with a thick He layer.  Then, on a timescale of
  $\sim 10$ Myr, gravitational wave radiation will again bring the
  systems into contact.  If the mass transfer is unstable and results
  in a merger and a catastrophic thermonuclear explosion is not
  triggered, it creates a remnant with a C/O-dominated envelope, but
  one still rich enough in He to support an R Corona Borealis-like
  shell burning phase.  We present evolutionary calculations of this
  phase and discuss its potential impact on the cooling of the remnant
  WD.
\end{abstract}


\section{Introduction}

Recently, \citet{Kupfer2020a, Kupfer2020b} described a new class of binaries consisting of a Roche lobe-filling hot subdwarf transferring mass to a
white dwarf (WD) companion.  The first of these objects, ZTF J213056.71+442046.5
(hereafter ZTF J2130+4420), is the most compact hot subdwarf binary
currently known with a period of $P=39.34$\,min.  Modeling of the
system yields a low-mass hot subdwarf donor with a mass
$M_{\rm sdOB}=0.337\pm0.015$\,\Msun\, and a white dwarf accretor with
a mass $M_{\rm WD}=0.545\pm0.020$\,\Msun.
The second of these objects, ZTF J205515.98+465106.5 (hereafter ZTF
J2055+4651), has a period of $P=56.35$\,min.  Modeling of the
system yields an He-sdOB mass of $M_{\rm sdOB}=0.41\pm0.04$\,\Msun\,
and a WD mass of $M_{\rm WD}=0.68\pm0.05$\,\Msun.
The timescales for these systems to merge (as WD-WD binaries) are
$\approx 17$ Myr and $\approx 30$ Myr respectively.

Similar binaries may realize a variety of final fates, depending on
the masses of the component objects and their post-common-envelope
orbital period.  \citet{Bauer2021} systematically map out the phases
of mass transfer in these systems, which can involve stable mass
transfer of both H- and He-rich material.  This can lead to outcomes
including the formation of an AM CVn system or the accumulation of a
thick He shell and the destruction of the system through a
thermonuclear double detonation \citep[e.g.,][]{Bauer2017b}.
Alternatively, the subdwarf donor may detach, leading to the formation
of a double WD binary with a C/O primary WD and low-mass hybrid He/C/O
WD. \citet{Zenati2019a} describe the formation and evolution of hybrid He/C/O WDs and characterize their properties and birth rates.

Gravitational wave radiation will again bring the systems into
contact.  The stability of the resulting mass transfer depends on the
spin-orbit coupling in the binary \citep[e.g.,][]{Marsh2004} and the
dissipative influence of material ejected in potential novae
\citep{Shen2015a}.  It is plausible that a significant fraction of the
systems experience unstable mass transfer and merge.

As discussed by \citet{Zenati2019a}, the significant He content of the He/C/O hybrid WD could have an
important role in the WD–WD merger.  In particular,
the He-rich secondary again offers the possibility of a
double-detonation thermonuclear explosion.
However, given the
relatively low mass of the primary C/O WD in the prototype systems
($\lesssim 0.7\,\Msun$), this outcome seems unlikely
\citep[e.g.,][]{Shen2018}.  If the merger does not lead to a
catastrophic explosion, it produces a remnant that will eventually
become a single massive WD.

In this paper, we illustrate this pathway by building a
simple model of the post-merger state and following its evolution.
Because of the significant amount of He present on the secondary WD,
the merger remnant first sets up a stable He-burning shell and spends
$\sim 10^5$ yr as a He-shell burning giant.  Once the He finishes
burning and/or is lost to stellar winds, the remnant contracts to a
compact configuration and then moves down the cooling track as a
massive WD.  Because of its merger origin, the composition profile can
be different than that of a WD from single star evolution, and we
discuss how this can affect the WD cooling through the production of
\neon[22].

A closely related scenario is the merger of a He-core WD with a
low-mass C/O-core WD.  This is thought to produce the R Coronae
Borealis (RCB) stars \citep[e.g.,][]{Webbink1984, Clayton2012} and
extreme He (EHe) stars \citep[e.g.,][]{Saio2002, Jeffery2008c}.  As we
outline the evolution, we compare and contrast with these objects.
The observed sample of RCB stars, with their measured surface abundances \citep[e.g.,][]{Asplund2000, Clayton2007}, motivates detailed studies of their evolution including particular attention to the nucleosynthesis \citep[e.g.,][]{Menon2013, Zhang2014b, Munson2021}.
  The calculations we present here are more schematic than the state-of-the-art RCB models.  The goal of this work is to consider a variation on the basic \citet{Webbink1984} scenario, in which the He-core WD secondary is instead a low mass C/O-core WD with a thick He layer, and illustrate that this may produce an unusual object akin to, but distinct from, the RCBs.

\section{Evolution to a Single WD}
\label{sec:single}

\subsection{Initial post-merger model}

Hydrodynamic simulations of double WD mergers show that immediately
after the merger (i.e., a few orbital periods after the tidal
disruption of the secondary WD), the merger remnant can be divided
into three main regions: the cold core of the primary, a hot
shock-heated envelope at the primary/secondary interface, and a
rotationally-supported disk of material from the secondary
\citep{Dan2014}.  Then, efficient angular momentum transport due to
magnetohydrodynamic processes internally redistributes angular
momentum on timescales much shorter than the thermal timescale of the
remnant \citep{Shen2012, Schwab2012}.  The remnant becomes primarily
thermally supported and approaches a quasi-spherical state allowing
its further evolution to be treated as a stellar evolution problem.

We construct and evolve stellar models using MESA r12778
\citep{Paxton2011, Paxton2013, Paxton2015, Paxton2018, Paxton2019}.
The initial merger model generation procedure is similar to that
described in \citet{Schwab2021}.  First, a high-entropy,
pure-helium model of the desired mass is constructed.  Next, its
composition and temperature/entropy profiles are slowly altered to
match specified target profiles (described subsequently).  This model
then forms the starting condition and is allowed to evolve without
further intervention.  The technical details of this procedure are
irrelevant given the illustrative nature of the calculation presented
here, but the MESA input files are available on Zenodo (\url{https://zenodo.org/record/5063047}).

We generate an initial condition that resembles a possible outcome from
the future evolution of ZTF J2130+4420 under the constraint that the pre-merger component
masses are among the set of \citet{Dan2014} WD-WD merger simulations.  Therefore, we
model $0.35\,\Msun$ C/O core WD with a thick He layer merging with
$0.55\,\Msun$ C/O WD.  In the \citet{Dan2014} simulation with this
total mass and mass ratio, they find
$M_{\rm core} \approx 0.4\,\Msun$, $M_{\rm env} \approx 0.3\,\Msun$,
and $M_{\rm disk} \approx 0.2\,\Msun$. (In this case, the
\citet{Dan2014} secondary was a He WD, but they demonstrate that
these quantities depend primarily on the mass ratio.)

We assign the core a cold temperature of $\unit[5\times10^7]{K}$.  The
peak temperature is $T_{\rm peak} \approx \unit[2\times10^8]{K}$ and
we choose a constant entropy for the envelope material that yields
this temperature.  The choice of the envelope entropy will affect the
early evolution, but after a thermal time, the entropy of the outer
layers is reset by the luminosity from below. Our model is
non-rotating.

In order to set the composition, we rely on a model of an
$\approx 0.6\,\Msun$ C/O WD primary generated from a $3.1\,\Msun$
single star and the $\approx 0.36\,\Msun$ WD secondary evolved from
the He-sdO model presented in \cite{Kupfer2020a}. In evolving the
He-sdO model into the WD stage, much of the H in the envelope burns
away to form a much thinner envelope than the $0.01\,\Msun$
envelope in the He-sdO phase presented in \cite{Kupfer2020a}.
The resulting secondary WD envelope has a total H mass of $\approx 8\times10^{-4}\,\Msun$.

\begin{figure}
  \centering
  \includegraphics[width=\columnwidth]{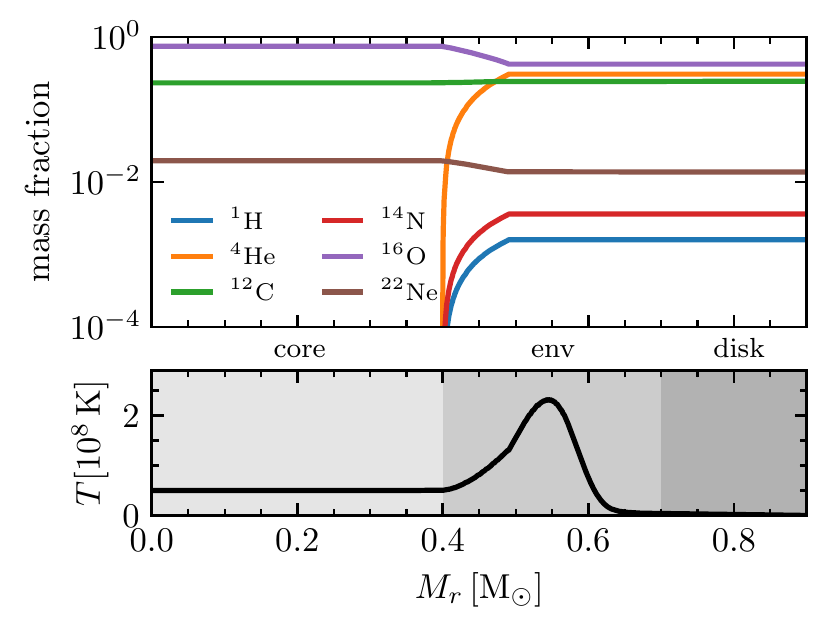}
  \caption{Initial condition representing the immediate post-merger
    state.  The top panel shows the mass fractions of key isotopes;
    the bottom panel shows the temperature.  The three key regions of
    the remnant are indicated/shaded. }
  \label{fig:ic}
\end{figure}

In the core region,
we set the composition to be the averaged mass fractions of the core
of the primary WD.  In the envelope/disk region, we use a set of
averaged mass fractions that reflects the result of uniformly mixing
the outer $\approx 0.15\,\Msun$ of the primary WD and the entire
secondary WD.  The composition transitions between these over a small
blend region located around $M_r = M_{\rm core}$.  This initial
condition is illustrated in Figure~\ref{fig:ic}.  The homogenization
of the outer layers is apparent. For example, the surface H layer from the
secondary is mixed throughout the envelope/disk region.

\subsection{Post-merger to WD cooling track}

Post-merger, the material near the temperature peak immediately begins
to burn.  The H is rapidly consumed (within days) and the He first
undergoes a flash and then sets up a steady shell burning structure
(in $\sim 10^2$ years).  The H burns through CNO reactions, converting
some of the \carbon[12] and \oxygen[16] to \nitrogen[14].  As He
burning begins, the reactions
$\nitrogen[14](\alpha, \gamma)\fluorine[18](\beta^+)\oxygen[18]$ and
$\oxygen[18](\alpha, \gamma)\neon[22]$ process some of the
pre-existing and newly-generated \nitrogen[14] into neutron-rich
isotopes.
The RCB stars exhibit large \oxygen[18]/\oxygen[16] ratios
\citep{Clayton2007}. Reproducing the \oxygen[18] and other observed
surface abundances provides constraints on the temperature and
lifetime of this hot region and the mixing processes that bring this
material to the surface \citep{Staff2012, Menon2013, Staff2018,
  Crawford2020}.

\begin{figure}
  \centering
  \includegraphics[width=\columnwidth]{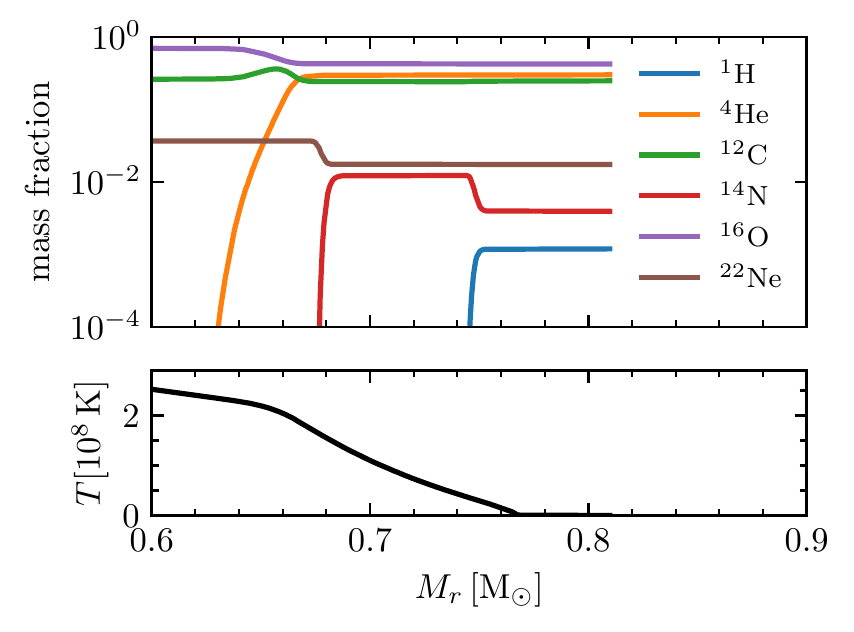}
  \caption{Model during the steady He-shell burning phase (after
    $\approx 170$ kyr of post-merger evolution).  The top panel shows
    the mass fractions of key isotopes; the bottom panel shows the
    temperature.}
  \label{fig:cn}
\end{figure}

While critical for setting the surface abundances, only
$\approx 0.1\,\Msun$ of material is processed in the hot, short-lived
interface region created in the merger.  The overall interior abundance
profile is also influenced by the evolution during the extended He-shell-burning phase.
Given the presence of a H mass fraction of $X_{\rm H} \sim 10^{-3}$ in
the envelope, during He shell burning the object also has a H burning
shell.  Because of the small H abundance, this shell is not
structurally or energetically important.  Because the \carbon[12]
abundance is much greater than the H abundance, the H is completely
consumed in converting the \carbon[12] to \nitrogen[14], increasing
the mass fraction of the latter isotope by an amount
$\Delta X_{\rm N} \approx 7 X_{\rm H}$.  Then, once this material
approaches the temperature of the He burning shell it is further
converted into \neon[22], leading to an enhancement in its mass
fraction of $\Delta X_{\rm Ne} \approx 11 X_{\rm
  H}$. Figure~\ref{fig:cn} shows a snapshot of the composition profile
during the giant phase that illustrates this process.

Because it is not directly observable, the evolution of the interior
profile has received less attention that the surface in RCB
modeling. \citet{Menon2013}, who start with
$X_{\rm H} \approx 10^{-2}$ motivated by the thick H envelope on a low
mass He WD secondary, achieve \neon[22] fractions
$X_{\rm Ne} \approx 10^{-1}$ in accordance with our above estimates
(see their Figures 5 and 6).

Exactly how much H is present at merger is unclear, and so the
predicted \neon[22] enhancements are correspondingly uncertain.  The H
layer masses at the formation of the secondary WD represent an upper
limit, as some of this H will be stably transferred in the lead up to
the merger \citep[e.g.,][]{Kaplan2012, Shen2015a}.  Empirically, the observed H
abundances in the RCB stars exhibit a large range
\citep{Asplund2000}.

\begin{figure}
  \centering
  \includegraphics[width=\columnwidth]{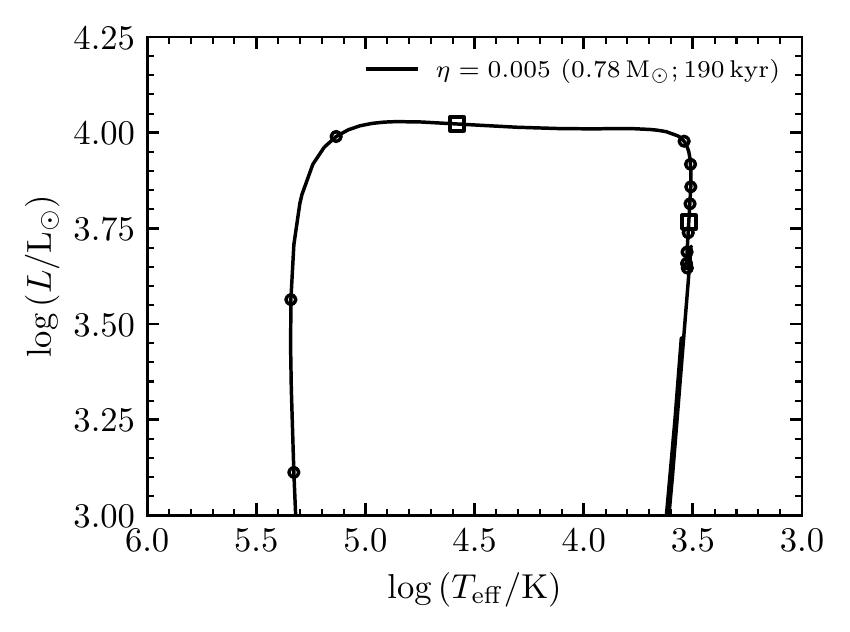}
  \caption{Evolution in the HR diagram.  Along the track, small
    circles are placed every 20 kyr and large squares every 100 kyr.
    The legend indicates the assumed mass loss parameter (using a
    \citealt{Bloecker1995a} wind prescription) with the final WD mass
    and giant phase ($\Teff < 10^4\,\rm K$) lifetime. }
  \label{fig:hr}
\end{figure}

Figure~\ref{fig:hr} shows the evolution in the HR diagram.  The remnant has a cool,
luminous giant phase lasting 190 kyr (analogous to the RCB phase) and
then has a shorter 10 kyr phase moving blueward (analogous to the EHe
phase).  This calculation adopts the \citet{Bloecker1995a} mass loss
prescription using the parameter $\eta = 0.005$.
While this evolution is broadly similar to RCBs/EHes, the
different compositions likely make these observationally distinct.
The critical difference from the RCB/EHe stars is that these objects
have C/O-dominated surfaces with some He, as opposed to He-dominated
surfaces with some C/O.  In this case, the abundances of the C/O core
of the secondary imply the outer material is expected to be
O-rich.  This will alter the chemistry of the cool outer layers.
The characteristic variability of the RCBs is due to the formation of
carbon dust, so these O-rich objects may have different
variability properties.

The lower total amount of He present when compared to an RCB star of
the same mass suggests a shorter He-shell-burning lifetime.  As in the
RCB case, a key determinant of the duration of this phase and of the
final mass of the WD is the mass loss rate \citep{Schwab2019c}.  The
mass loss rates are theoretically uncertain.  In AGB stars, the
chemistry influences the wind properties \citep[e.g.,][]{Hofner2018},
though the extent to which AGB mass loss rates are appropriate to
apply in this circumstance is unclear.  If, for example, a lack of
carbon dust formation reduces the mass loss rate relative to mass loss
models calibrated on the C-rich RCB stars, this would lead to longer
predicted lifetimes.

If, as a crude estimate, we say that the formation rate of these
objects is a factor of $\sim 10$ lower than for RCB stars due to the
requirement for more massive stellar progenitors and the lifetimes are
a factor of $\sim 3$ shorter, then this would suggest $\sim 5$
galactic objects in this phase relative to the $\approx 150$ known RCB
stars \citep[e.g.,][]{Tisserand2020}. \citet{Bauer2021} estimate that
the Galaxy contains at least $\sim 10^4$ sdB+WD binaries with
orbital periods shorter than 1–2 hours, with lifetimes in this phase
of $\sim 100$~Myr.  If an order-unity fraction of these objects have an
outcome like that discussed in this paper, with remnant lifetimes of
$\sim 100$~kyr, then this too plausibly suggests the Galaxy contains
at least $\sim 10$ merger remnants currently in this phase.
On the other hand, the population synthesis results of
  \citet{Zenati2019a} suggest that WD mergers involving a hybrid
  He/C/O WD could occur at a rate of $\sim 10^{-3}\,\rm yr^{-1}$ in
  our Galaxy. With a remnant lifetime of $\sim 100$~kyr, this would
  predict as many as $\sim 100$ galactic objects currently in this
  phase.

\begin{figure}
  \centering
  \includegraphics[width=\columnwidth]{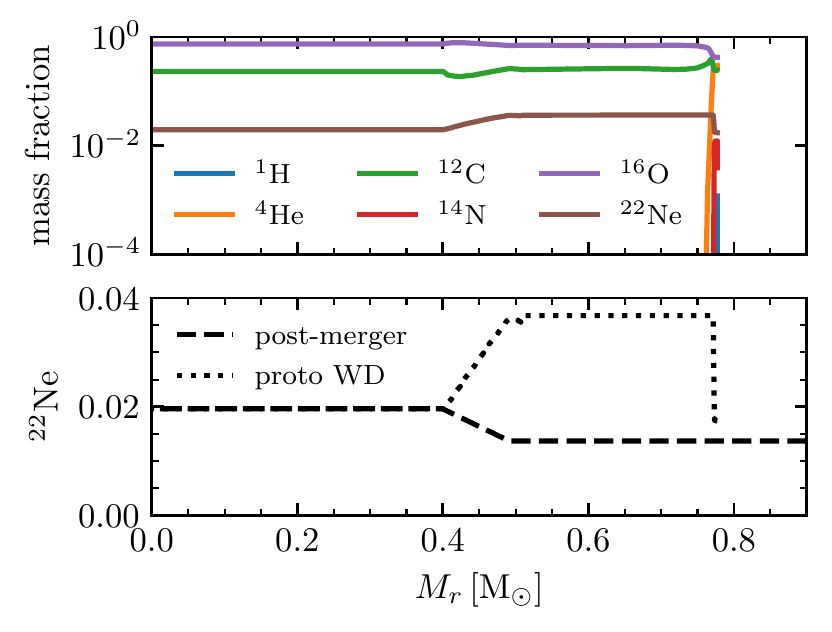}
  \caption{Composition profile when the object is a hot proto-WD.
    The top panel shows the mass fractions of key isotopes; the bottom
    panel illustrates the change in the \neon[22] abundance profile
    between this time and the post-merger initial condition shown in
    Figure~\ref{fig:ic}.}
  \label{fig:protowd}
\end{figure}

\subsection{On the WD cooling track}

Figure~\ref{fig:protowd} shows the interior composition profile at the
time when the object is a hot proto-WD.  The object now has a C/O
core, and mass loss has reduced the total mass by
$\approx 0.1\,\Msun$.  As shown in the bottom panel, the material
outside the core has experienced an enhancement in \neon[22] mass
fraction up to roughly twice its initial value.  During the early
phases of the cooling, convection and thermohaline mixing driven by
the electron fraction gradient can redistribute this neutron-rich
material deeper into the interior \citep[e.g.,][]{Brooks2017a, Schwab2019b}.

Recently, \citet{Camisassa2021} invoked mergers to produce WDs with high
\neon[22] mass fractions ($\approx 0.06$) and correspondingly produce significant delays in the
WD cooling due to sedimentation heating.  Our calculations do find 
\neon[22] enhancement as a result of the merger, with
the expected mass fraction effectively determined by
the (uncertain) mass fraction of H present in the mixed, post-merger envelope.
It is important to note that this level of enhancement is not uniform, but
only expected in material processed through the H- and He-burning shells present
in the remnant.
Thus, the total amount of $\neon[22]$
enhancement will also depend on the mass loss rate of the remnant as this sets the
fraction of the envelope material incorporated on the final WD.  Given
their assumed mass loss rates, many RCB models find most of the
envelope is lost
\citep[e.g.,][]{Zhang2014b,Lauer2019,Crawford2020,Munson2021}, which
can then limit the total \neon[22] enhancement.

\section{Conclusions}

We have outlined one future evolutionary pathway for objects in the
new class of hot subdwarf - WD binaries described by
\citet{Kupfer2020a, Kupfer2020b}.  After evolving to detached WD-WD
binaries, gravitational waves will cause these systems to inspiral and
their low primary WD masses suggest that they may not suffer a
catastrophic thermonuclear explosion at merger.  This implies the
formation of a long-lived remnant and that their final fate will be a
single WD.

The properties of these WD-WD binaries resemble the He WD - C/O WD binaries
thought to form the RCB stars and so suggest a similar evolutionary
trajectory, though with a few key differences.  Because the secondary
WD (descended from the hot subdwarf) has already undergone core He
burning, the total He supply is reduced (compared to a merger
involving a He WD secondary).  However, enough He still exists to
allow for the formation of a He-shell-burning powered giant star with
a C/O-dominated envelope.  This giant phase is similar to that of the
RCB stars, but with a different envelope composition.  Notably, the
expectation is that the envelope is O-rich, which may lead to
different mass loss and photometric variability properties.  In turn,
that implies these objects might not be easily detected using
techniques optimized for the RCB stars (and their C-rich dust
formation events).

The nucleosynthesis that occurs in the post-merger envelope provides a
pathway for the enhancement of \neon[22].  The amount of \neon[22]
produced depends on the amount of H incorporated in the
post-merger envelope and on the fraction of that envelope that ends up
being processed and retained (i.e., not lost to stellar winds).
Through their influence on the energy released via gravitational
sedimentation and phase separation, these composition differences may
manifest as differences in the detailed cooling properties of single
WDs produced in mergers.  However, additional progress modeling the
pre-and-post merger evolution will be required reliably assess the
magnitude of these effects.

\begin{acknowledgements}

We thank Thomas Kupfer and Ken J. Shen for helpful conversations.  
JS is supported by the A.F. Morrison Fellowship in Lick Observatory,
the National Science Foundation through grant ACI-1663688, and via
support for program number HST-GO-15864.005-A provided through a grant
from the STScI under NASA contract NAS5-26555.
This work used the Extreme Science and Engineering Discovery Environment \citep[XSEDE;][]{XSEDE2014}, which is supported by National Science Foundation grant No.~ACI-1548562, specifically comet at the SDSC through allocation TG-AST180050.
This research has made use of NASA's Astrophysics Data System Bibliographic Services.

\end{acknowledgements}

\software{
\MESA\ \citep{Paxton2011, Paxton2013, Paxton2015, Paxton2018, Paxton2019},
\texttt{MESASDK} 20.3.1 \citep{mesasdk}, 
\texttt{matplotlib} \citep{hunter2007}, 
\texttt{NumPy} \citep{walt2011},
\texttt{py\_mesa\_reader} \citep{pmr},
\texttt{MesaScript} \citep{MesaScript}
}

\appendix

\section{MESA r12778 Input Physics}

  The MESA EOS is a blend of the OPAL \citet{Rogers2002}, SCVH
  \citet{Saumon1995}, PTEH \citet{Pols1995}, HELM \citet{Timmes2000},
  and PC \citet{Potekhin2010} EOSes.
  
  Radiative opacities are primarily from OPAL \citep{Iglesias1993,
  Iglesias1996}, with low-temperature data from \citet{Ferguson2005}
  and the high-temperature, Compton-scattering dominated regime by
  \citet{Buchler1976}.  Electron conduction opacities are from
  \citet{Cassisi2007}.
  
  Nuclear reaction rates are from JINA REACLIB \citep{Cyburt2010} plus
  additional tabulated weak reaction rates \citep{Fuller1985, Oda1994,
  Langanke2000}. Screening is included via the prescription of \citet{Chugunov2007}.
  Thermal neutrino loss rates are from \citet{Itoh1996}.

\clearpage
\bibliography{He_sdO_WD_fates.bib}

\end{document}